# Enhancing the response of NH$_3$ graphene-sensors by using devices with different graphene-substrate distances

A. R. Cadore,[1,*)] E. Mania,[1] A. B. Alencar,[2] N. P. Rezende,[1] S. de Oliveira,[1] K. Watanabe,[3] T. Taniguchi,[3] H. Chacham,[1] L. C. Campos,[1] R. G. Lacerda[1,*)]

[1]*Departamento de Física, Universidade Federal de Minas Gerais, Belo Horizonte, 30123-970, Brasil*

[2]*Instituto de Engenharia, Ciência e Tecnologia, Universidade Federal dos Vales do Jequitinhonha e Mucuri, Janaúba, 39440-000, Brasil*

[3]*National Institute for Materials Science, Namiki, 305-0044, Japan*

[*)]*Electronic mail: alissoncadore@gmail.com; rlacerda@fisica.ufmg.br*

Graphene (G) is a two-dimensional material with exceptional sensing properties. In general, graphene gas sensors are produced in field effect transistor configuration on several substrates. The role of the substrates on the sensor characteristics has not yet been entirely established. To provide further insight on the interaction between ammonia molecules (NH$_3$) and graphene devices, we report experimental and theoretical studies of NH$_3$ graphene sensors with graphene supported on three substrates: SiO$_2$, talc and hexagonal boron nitride (hBN). Our results indicate that the charge transfer from NH$_3$ to graphene depends not only on extrinsic parameters like temperature and gas concentration, but also on the average distance between the graphene sheet and the substrate. We find that the average distance between graphene and hBN crystals is the smallest among the three substrates, and that graphene-ammonia gas sensors based on a G/hBN heterostructure exhibit the fastest recovery times for NH$_3$ exposure and are slightly affected by wet or dry air environment. Moreover, the dependence of graphene-ammonia sensors on different substrates indicates that graphene sensors exhibit two different adsorption processes for NH$_3$ molecules: one at the top of the graphene surface and another at its bottom side close to the substrate. Therefore, our findings show that substrate engineering is crucial to the development of graphene-based gas sensors and indicate additional routes for faster sensors.

**Keywords**: ammonia detection, graphene, gas sensor, substrate engineering

1.  **INTRODUCTION**

The discovery of two-dimensional (2D) materials [1] like graphene (G), semiconductor transition metal dichalcogenides and 2D insulators have brought exciting predictions for sensing devices due to their electrical, thermal and surface properties [2–4]. For instance, graphene-based sensing systems have been under intensive investigations since its whole area is capable of interacting with the surrounding gas [5–8], making graphene an ultrasensitive material for gas detection [9]. However, the mechanism of interaction between graphene and the adsorbing molecules is particularly dependent of each target chemical species [3–6,10–12]. Among the diversity of existing target gases, ammonia ($NH_3$) has been widely studied for graphene-based sensors [13–23] due to its great significance for industrial applications [24]. Theoretical works show that charge transferred from $NH_3$ to graphene is dependent on the orientation of the $NH_3$ molecules [25,26]. Conversely, experimental works indicate that other parameters like electric field [13,19], doping from the substrates [14], chemical functionalization [20] and surface decoration [23] also affect ammonia detection. Besides, recent works [27–30] demonstrated that substrate engineering such as using microporous surface is able to improve considerably graphene sensing properties in comparison to the flat one. However, the details about the influence of the substrate, temperatures and $NH_3$ concentrations has not been completely established, requiring further theoretical and experimental investigations.

In this work we compare the performance of graphene-based sensors for ammonia gas supported on three different dielectric substrates ($SiO_2$, talc and hexagonal boron nitride (hBN)). We choose $SiO_2$ for being the most common substrate used in graphene devices [1] and the other two for being atomically flat substrates that produces graphene devices with different qualities. For example, hBN is the platform that enables graphene devices to reach its highest carrier mobility [31], whereas talc ($Mg_3Si_4O_{10}(OH)_2$) is a natural silicate dielectric where graphene devices display the best electronic quality on an oxide substrate [32]. Our investigation indicates that the gas sensing characteristics depends mainly on the average distance between the graphene sheet and the substrate. In sensors with graphene on top of hBN, which have the smallest G/substrate distance (0.5 nm), the sensor exhibits the fastest recovering time under $NH_3$ exposure and are slightly affected by air environment. On the other hand, for graphene on top of $SiO_2$, where the graphene-substrate distance is approximately 1.0 nm, the diffusion of $NH_3$ in between graphene and $SiO_2$ leads to interaction at two sides of graphene, resulting in a larger electron charge transfer. This observation agrees with our prediction by density functional theory (DFT). The dependence of the properties of $NH_3$ graphene-sensor with the substrate type also confirms that graphene-based sensors exhibit two different adsorption modes: one at the graphene top side and another at its bottom side. Furthermore, our findings show that typical sensor properties such as resistance response and recovering time of a $NH_3$ graphene-sensor can be modified by choosing a desirable substrate, revealing that substrate engineering may be crucial to the development of graphene-based sensors and electronic devices.

## 2. EXPERIMENTAL METHODS

In order to investigate the influence of the separation between graphene and substrates on $NH_3$ graphene sensors, we prepare graphene sensor on three substrates: $SiO_2$, hBN and talc. Monolayer graphene, hBN and talc are prepared by the scotch tape method [1,33] and transferred atop of desired substrates by using a stacking method [34]. To attain the best surface flatness, hBN and talc crystals were prepared with thickness in between 15 nm and 30 nm, which provides inert, very flat and clean surfaces [31,32]. Indeed, any thickness higher than 15 nm did not produce any appreciable changes in device and sensing performance. On the other hand, it is known that for thickness smaller than 10 nm, charge tunneling through the dielectric may happen [35], which would lead to gate leakage current.

The graphene devices on top of $SiO_2$/Si (G/$SiO_2$) are obtained by mechanical exfoliation directly to the $SiO_2$ substrate. Graphene devices on top of hBN (or talc) substrates are fabricated as follows: 10-25 nm graphite thick is exfoliated on $SiO_2$, to be used as a back-gate electrode; then, a selected hBN (or talc) flake is transferred first, followed by graphene on top of it, forming a graphene/hBN/graphite (G/hBN) or graphene/talc/graphite (G/talc) heterostructures. The use of graphite as a back-gate electrode is desirable in order to avoid charging effects at the hBN/$SiO_2$ (or talc/$SiO_2$) interfaces [36]. After each material transference, the devices are submitted to a heat cleaning at 623 K with continuous flow of Ar:$H_2$ (300:700 sccm) for 3.5 h to eliminate organic remains.

In total, we examined three G/hBN, two G/talc and three G/$SiO_2$ devices, which all produced consistent results. Graphene, talc and hBN flakes are characterized by atomic force microscopy (AFM) to assess flatness and cleanliness, and by optical investigation to identify monolayer graphene [37]. Electron-beam lithography and oxygen plasma etching are used to define the graphene geometry, and thermal evaporation of Cr/Au is used to fabricate the contacts (1/40 nm). The dimensions of the graphene channel in all the devices are approximately 1 µm:1.8 µm (length : width). Finally, to remove polymer residues remaining from the lithography processes, the devices are submitted to a final heat cleaning process.

Each graphene device is then mounted on a ceramic chip carrier and placed in a socket fixed inside a homemade gas handling system tube equipped with heater, electrical connections (power supply and computer) and gases (Ar, $N_2$, dry air and $NH_3$) mass flow controllers (MFC) as shown in Fig. 1 (a). Thermal annealing of the devices is performed in situ by heating to 473 K and keeping during overnight in pure $N_2$ atmosphere to drive out all undesirable gases [38]. Once the annealing procedure is ceased and graphene reaches a fully degassed state, the samples can cool down to the temperature of interest and are ready to be exposed to anhydrous $NH_3$ (99.99 %, $H_2O$ < 1ppm). Note that we consider as fully degassed state the situation where there is no further evolution of the charge neutrality point (CNP) position of the graphene devices under a certain temperature. The electronic measurements of all devices are performed in a four-terminal configuration using lock-in technique at 17 Hz with a current bias of 1 µA applied between source and drain electrodes, as illustrated in Fig. 1 (b). Moreover, their sensors response is tested at different operation temperatures, ranging from 300 K up to 450 K and ammonia concentration ranging from 2.5 % up to 20 %. Note that to attain such $NH_3$ concentrations, we diluted 100 % $NH_3$ gas with different carrier gases.

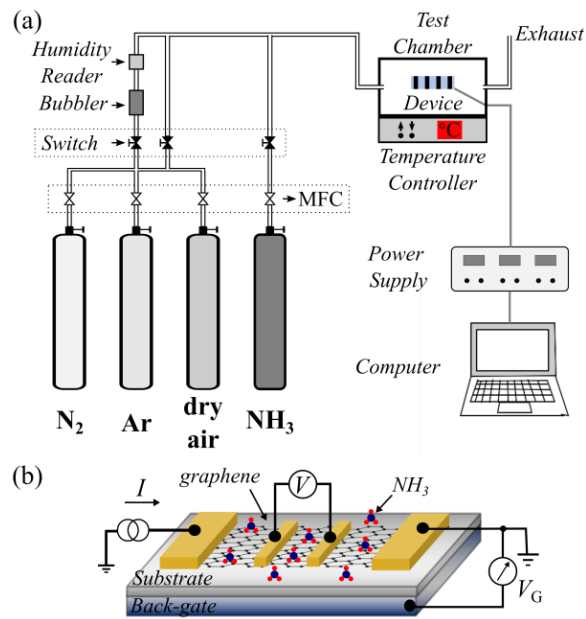

Figure 1: (a) Schematic diagram of the experimental setup which includes a test chamber furnace equipped with a heater, electrical connections (power supply and computer), switches, gases ($N_2$, Ar, dry air and $NH_3$), mass flow controllers (MFC), a bubbler and humidity reader. (b) Schematic of a back-gated graphene device on top of a general substrate during $NH_3$ exposure with also a representation of the current-biased measurement setup.

In this work the resistance response is taken by the average value of all exposure runs $\left(Response = \frac{1}{N}\sum_1^N\left((R_{Dg} - R_{NH_3})/R_{Dg}\right) \times 100\%\right)$, where $N$ is the number of pulses repeated for each ammonia concentration; $R_{Dg}$ is the resistance under pure diluting gas (Nitrogen ($N_2$), Argon (Ar) or dry air); $R_{NH_3}$ is the resistance under the ammonia exposure. The measurements presented here were performed using different diluting gas (ultrahigh pure $N_2$, Ar and dry air), which all demonstrated similar results, indicating that all gases are inert for ammonia detection and do not change the interaction between both sides of the graphene layer and $NH_3$ molecules Additionally, we also investigated in Fig. 6 the effect of dry air and relative humidity on the NH$_3$ response properties for the G/hBN sensors. More details on the influence of dry air and humidity on sensing response for other devices are presented in supplementary material.

### 3. RESULTS AND DISCUSSION

Initially, in order to study the influence of the G/substrate distance on the sensing characteristics, careful analyses of the graphene-surface were taken using AFM to exclude any effect of residues, wrinkles and defects that may be created during the fabrication process. The height distribution (surface roughness) of G/hBN, G/talc and G/SiO$_2$ devices are compared in Fig. 2 (a). The figure depicts that the results for G/hBN and G/talc are similar, and both are narrower than in G/SiO$_2$. The root mean square height values ($h$RMS) obtained from the Gaussian fits (solid lines in Fig. 2(a)) at each surface with the graphene layer are: $h$RMS ~ 0.10 nm for G/hBN, $h$RMS ~ 0.15 nm for G/talc and $h$RMS ~ 0.20 nm for G/SiO$_2$, while without the graphene are: $h$RMS ~ 0.08 nm for hBN, $h$RMS ~ 0.16 nm for talc and $h$RMS ~ 0.26 nm for G/SiO$_2$. Such values are in agreement with previous works [31,32,39], indicating that the graphene surface is clean and free of polymer residues.

Besides that, measurements of the G/substrate distance were performed to further verify the performance of graphene sensor and our initial hypothesis of preparing graphene devices with different G/substrate separation. The topography profiles indicating the G/substrate distance (thickness) for each sample can be observed in Fig. 2 (b). One can note that G/hBN device (red curve) have the smallest G/substrate distance of about 0.5 nm, while G/talc sample (blue curve) and G/SiO$_2$ device (black curve) have larger ones of about 0.8 nm and 1 nm, respectively. The G/substrate distance for our G/SiO$_2$ devices is slightly higher than the values found in the literature for a graphene layer on top of SiO$_2$ (0.9 nm) [40,41]. Nevertheless, it is important to mention that in our G/SiO$_2$ samples, the SiO$_2$/Si substrates are etched for 1 min in oxygen plasma before graphene exfoliation, and it is well-known that such process changes the oxide surface increasing its surface roughness [42,43], which in turn also increases the G/substrate distance. In addition, Fig. 2 (c) shows AFM images of ordinary G/hBN, G/talc and G/SiO$_2$ devices which are free of bubbles, wrinkles, or polymer residues on the graphene channel region. All AFM measurements were performed after thermal annealing at 623 K with constant flow of Ar:H$_2$ and any residual remains from the fabrication process was swept off from the graphene layer using an AFM tip operated in contact mode [44]. The studies presented above are important in order to attain that all G/SiO$_2$, G/hBN and G/talc devices have a flat and clean graphene surface, eliminating any false contribution to sensing response [40].

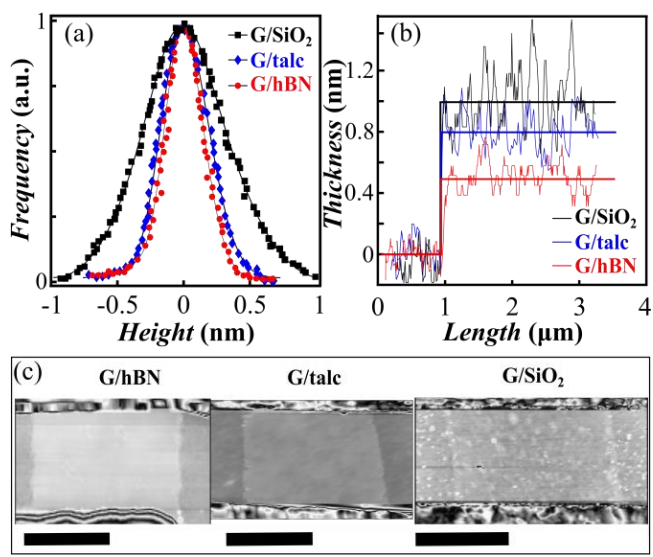

Figure 2: (a) Height histogram distributions (surface roughness) measured by AFM for G/SiO$_2$ (black data), G/talc (blue data) and G/hBN (red data). Solid lines are Gaussian fits to the height distribution. (b) AFM topography profile of graphene supported on all substrates showing the graphene-substrate distance (thickness) on SiO$_2$ (black data), talc (blue data) and hBN (red data). (c) AFM false-color image of the graphene layer on top of hBN, talc and SiO$_2$ substrates. In (c) the scale bars are 1µm.

Subsequent, we investigate the gas sensing characteristics of the graphene devices at 300 K, exposing them to 10 % NH$_3$ gas diluted in ultrahigh pure N$_2$ for 60 min. In all devices, the charge neutrality point (CNP) shifts towards negative gate voltages confirming the negative charge transfer to graphene devices when exposed to NH$_3$ (see Fig. S1 of supplementary material). Also, the electron donor character for the NH$_3$ molecules is consistent with our DFT calculations (see supplementary material), and it is in agreement with previous theoretical and experimental studies [14–16,19]. To estimate the amount of charge per unit of area transferred from ammonia to graphene, we perform experiments sweeping the gate voltage (see Fig. S1 in the supplementary material). More precisely, from the shift of the CNP, before and after been exposed to NH$_3$ ($\Delta V^{CNP}$), we calculate the absolute value of the change of graphene charge density ($\Delta n$) by the formula: $\Delta n = (\varepsilon \varepsilon_0 \Delta V^{CNP})/ed$ [13,45]. In this equation, $\varepsilon$ is the dielectric constant of the substrate (SiO$_2$ = talc = hBN = 3.9), $\varepsilon_0$ is the vacuum permittivity, $d$ is the dielectric thickness (SiO$_2$ = 285 nm, talc = 18 nm or hBN = 20 nm), and $e$ is the electron charge (more details see supplementary material).

In Fig. 3 we plot $\Delta n$ as a function of time under 10 % of NH$_3$ for all devices at 300 K. Clearly, the amount of charge transferred to graphene changes, depending on the underling substrate. In G/SiO$_2$ sensors (Fig. 3 (a)) the amount of charge transferred to graphene is approximately 1.4 times larger than $\Delta n$ in G/talc devices (Fig. 3 (b)), and about 2 times larger than the doping detected in G/hBN heterostructures (Fig. 3 (c)). A larger charge transfer in G/SiO$_2$ devices is understandable, since we can imagine that NH$_3$ molecules can interact on both sides of graphene (top side and bottom side). Here we argue that SiO$_2$ roughness and the larger average separation between graphene and SiO$_2$ substrates allow diffusion of NH$_3$ molecules under the graphene sheet. In addition, in graphene G/SiO$_2$ devices we observe a slighter faster charge transfer saturation as it will be discussed later. Another important property of ammonia-graphene sensors is the recovering time, which means the time necessary to recover the initial conditions after stop interacting with NH$_3$. To measure the recovering time, we stop flowing NH$_3$ in the test chamber and we keep the devices under a constant flow of (500 sccm) of pure N$_2$. None of the devices returned to its original condition even after 2 h under N$_2$ flow. To accelerate the recovery process, the devices are annealed up to 473 K for 1 hour. This process is found to be sufficient to restore the initials values (before NH$_3$ exposure), and similar thermal treatment has been made for G/mica devices [14]. After reaching such condition, the samples are allowed to cool down to temperature of interest before another set of measurements.

As we qualitatively argue, the top side and bottom side of the graphene sheet are not equally accessible to NH$_3$ molecules. A faster adsorption process is expected at the top side of graphene, and a slower process at the bottom side. The latter process is kinetically hindered by slow gas diffusion at the interface between the graphene and substrate. The drive of diffusion is normally correlated with the distance between both materials and the ability of the graphene sheet to conform on the substrate topography [10,15,45]. For a better understanding of these processes, characteristic time constants related to the interaction between ammonia and graphene during the adsorption process were inspected using a double exponential function to fit out the experimental data as: $V_{\text{shift}}(t) = V_\infty + V_1\exp(-t/\xi_1) + V_2\exp(-t/\xi_2)$, where $t$ is the time, $\xi_1$ and $\xi_2$ are characteristic time constants, and $V_\infty$ is the steady state position of the CNP after long-term adsorption [15,45]. All graphene devices are prepared with the same area, so it is expected that molecules will be similarly adsorbed on the top surface for all devices, although the bottom sites will be filled differently by diffusion of NH$_3$ into the interfaces, depending of each substrate.

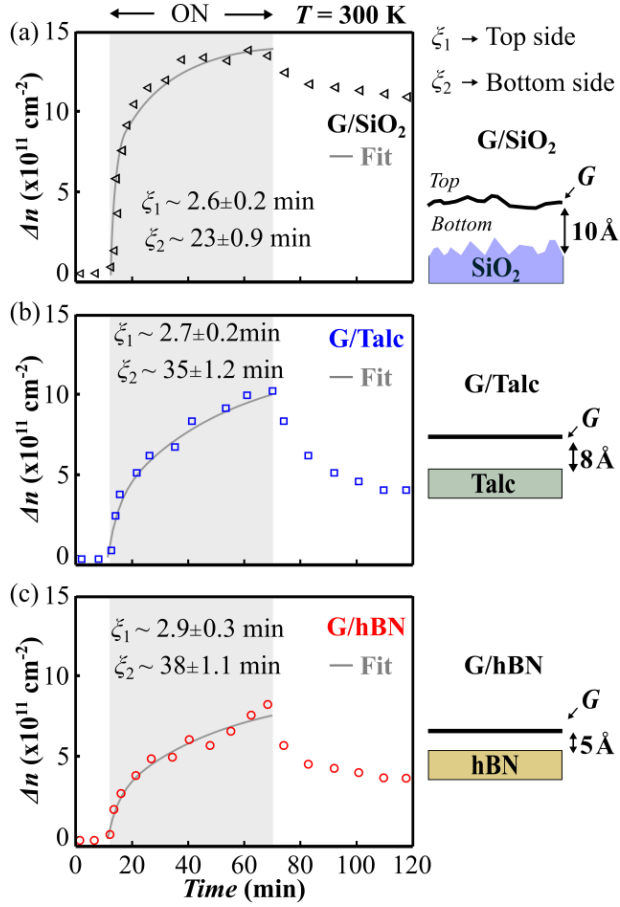

Figure 3: Absolute values of the change of the graphene charge density as a function of time. (a) Graphene on SiO$_2$; (b) Graphene on talc; and (c) Graphene on hBN substrates at 300 K. All measurements are performed under 10 % of NH$_3$ in N$_2$ as diluting gas. The gray regions show the time interval when the ammonia gas is turned ON into the system. Exponential double fittings of $\Delta n$ are shown by gray solid lines in the figures, and the respective time constants for each substrate are also shown. The right panels illustrate the average difference between the graphene sheet (G) and the substrates measured by AFM in Fig 2 (b). Note that $\xi_1$ is associated to the graphene top side, while $\xi_2$ is related to its bottom side, and for clarity, we show in figures (a-c) only few experimental data points, but the time in between two measurements is 1 min.

As seen in Fig. 3, the double-exponential equation provides a good description of the experimental observation for all substrates analyzed. The values of the times constants obtained are $\xi_1^{SiO_2} = 2.6 \pm 0.2$ min and $\xi_2^{SiO_2} = 23 \pm 0.9$ min ($\xi_1^{SiO_2}/\xi_2^{SiO_2} \approx 0.11$); $\xi_1^{talc} = 2.7 \pm 0.2$ min and $\xi_2^{talc} = 35 \pm 1.2$ min ($\xi_1^{talc}/\xi_2^{talc} \approx 0.08$); while $\xi_1^{hBN} = 2.9 \pm 0.3$ min and $\xi_2^{hBN} = 38 \pm 1.1$ min ($\xi_1^{hBN}/\xi_2^{hBN} \approx 0.08$) for SiO$_2$, talc and hBN, respectively. Note that $\xi_1$ values are similar for all sensors, supporting the idea that the fast process occurs at the top surface with a similar speed for all devices. The small difference between them can be caused, for instance, by charge inhomogeneity or polymer residues [40]. However, the $\xi_2$ constants are significantly different for each substrate. One can note that the value of $\xi_2^{SiO_2}$ is almost half of the $\xi_2^{talc}$ or $\xi_2^{hBN}$. We associate such difference to the diffusion of the molecules between graphene sheet and substrate, hence molecules would diffuse faster and easier between graphene on SiO$_2$ than on talc or hBN. The analysis presented here reinforces our proposal that the substrate plays an important role on the diffusion and detection of NH$_3$ molecules by graphene-based sensors.

Moreover, beyond the understanding of the influence of the two-adsorption kinetics involving top and bottom sides of the graphene sheet, and to develop a practical gas sensor, one must consider other important characteristics such as repeatability, recovery and gas response. Therefore, we provided an investigation of these aspects by performing measurements during pulses (1 min) of ammonia exposure at different concentration (from 2.5 % up to 20 %). We could not reduce the concentration below 2.5 % due to the limitations of our experimental setup, and we did not measure at higher concentration for safety reasons. Since we are interested in reducing the influence of graphene bottom side, we choose an exposing time exposure in such way that *exposing time* $\ll \xi_2$.

Fig. 4 (a) shows the time evolution of the resistance response for pulses (1 min ON and 5 min OFF) of different ammonia concentrations at 300 K. All systems show a good repeatability for three runs under the same conditions, and G/talc and G/hBN sensors recover faster than G/SiO$_2$. This characteristic will be discussed in detail later. Moreover, G/SiO$_2$ device exhibits a larger baseline (i.e., the difference in heights between the measured curve and the dashed line that is expected for pure diluting gas), indicating that the ammonia molecules are trapped at the interface between graphene and SiO$_2$ even for short exposure times. This indicates that G/SiO$_2$ sensors exhibit problems of poisoning that may diminish the ammonia detection after several cycles of exposure. For such analysis, we measure the change of the device resistance as a function of time at fixed gate voltage ($V_G$) such as: $V_G$ = - 60 V for G/SiO$_2$, $V_G$ = 3 V for G/Talc and $V_G$ = -3 V for G/hBN devices. Note that for all devices, we choose specific gate voltages to set an electronic conduction in the graphene channel with hole-type charge carriers, and that such electrostatic condition is far away from the CNP. Such choice was performed in order to ensure that the graphene would still be with hole-type as charge carries under NH$_3$ exposure (see supplementary material). Thus, when the NH$_3$ gas is turned ON, the adsorbed molecules transfer electrons to the graphene channel, which decreases the density of holes and hence increases the graphene channel resistance as expected [14,16,46] and discussed by our DFT calculations (see supplementary material).

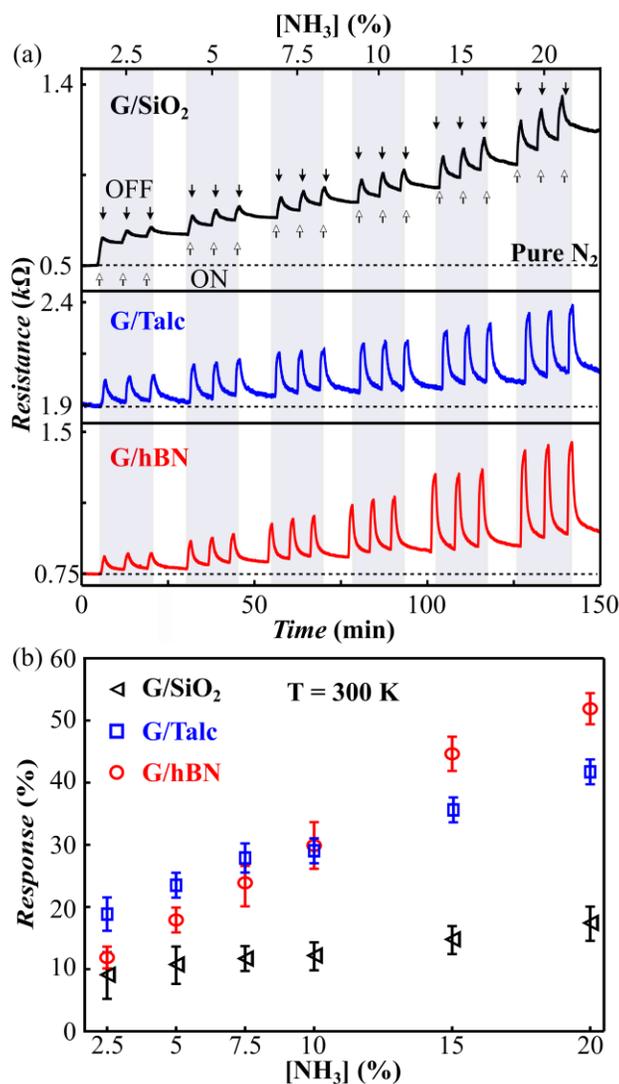

Figure 4: (a) Graphene resistance as a function of time for pulses - 1 min ON (arrows up) and 5 min OFF (arrows down) - of different NH$_3$ concentrations in N$_2$ as diluting gas at 300 K for G/SiO$_2$ (black curve), G/talc (blue curve) and G/hBN (red curve) devices. The dashed lines in (a) indicate the baseline expected solely for pure N$_2$ gas. (b) Resistance response as a function of ammonia concentration for pulses of different concentrations for all substrates analyzed.

In Fig. 4 (b) we present the resistance response for the pulses as a function of the ammonia concentration. From these measurements one observes that the sensors with small distance separation (G/talc and G/hBN) show a significant higher resistance variation to ammonia molecules than G/SiO$_2$. Therefore, comparing such results, our findings confirm that the distance between graphene and substrate plays an important role on the graphene sensor characteristics for NH$_3$ detection. Our results are in agreement with Aziza and co-workers that the G/substrate distance still play a significant role in the ammonia detection even for ppm levels [14]. In this work, the authors observed that graphene on mica substrate shows higher NH$_3$ sensitivity than graphene on SiO$_2$/Si substrates down to 20 ppm. The authors discussed the superior response for G/mica sensors

This is the authors' version pre-peer reviewed of the following article:
Alisson R. Cadore et. al., "Enhancing the response of $NH_3$ graphene-sensors by using devices with different graphene substrate distances".
*Sensors and Actuators B: Chemical (2018), DOI: 10.1016/j.snb.2018.03.164*based on the hydrophobicity and the electron acceptor behavior of the mica substrate, leaving graphene as a conductor with hole-type as charge carriers. Similar process are also observed for our G/talc devices (see Fig. S1c) and discussed in detail in our previous work [32]. Nevertheless, it is important to point out that mica has an atomically flat surface [39,47,48], and previous works demonstrated that the G/mica distance is smaller than G/$SiO_2$, and comparable to G/hBN, reaching values around 0.5 nm. Therefore, such results are completely consistent with our analyses and can be explained based on the G/substrate distance (G/mica smaller than G/$SiO_2$) rather than by the initial doping and substrate hydrophobicity, corroborating with our findings discussed here.

Besides the mentioned changes in graphene resistance and charge transfer, the ammonia gas also causes changes in graphene electronic mobility (see supplemental material), reducing both electron and hole maximum mobility. This suggests that the adsorbed ammonia molecules is also acting as charge scattering impurities [49,50]. Another aspect that must be pointed out is that by analyzing Figs. 3 and 4, one would expect that a larger charge transfer would cause also a superior resistance response. On the contrary, we observe that for the G/$SiO_2$ device (with larger charge transference) there is a smaller resistance response. The explanation for this process is still unclear, but our data supports the evidence that charge scattering (due to the presence of $NH_3$ molecules) affects more effectively high-quality devices, decreasing charge mobility. We believe that a more detail studied should be performed to clarify the influence of these two different mechanisms (charge transferring and changes on charge mobility). At this stage such analyzes is out of the scope of this work.

We now investigate the recovering process of the sensors. In this analysis we consider times under exposure to $NH_3$ of either 30 min or 1 min. In Fig. 5 we present the time evolution of resistance for a fixed ammonia concentration (10 %) at 300 K (similar characteristics are obtained for the other concentrations). Note that the recovering of the sensors is not ideal (100 %). For long time exposure (30 min), the G/$SiO_2$ sensor recovers around 25 %, while G/talc and G/hBN are able to recover up to 70 % and 60 %, respectively. Such low recovering time (> 1 h) to reach the baseline value has already been discussed for $NH_3$ molecules desorption, suggesting that resetting the sensor requires further efforts such as UV light irradiation [21], thermal or vacuum annealing [14,45] and water-vapor-enriched air [16]. However, for small time (1 min) under exposure to $NH_3$, the recovering enhances considerably: G/$SiO_2$ increases up to 50 %, while G/talc and G/hBN goes up to 85 %, close to the ideal case. Such enhancement in recovering may be caused by the lack of considerable diffusion of $NH_3$ at the graphene sheet bottom side.

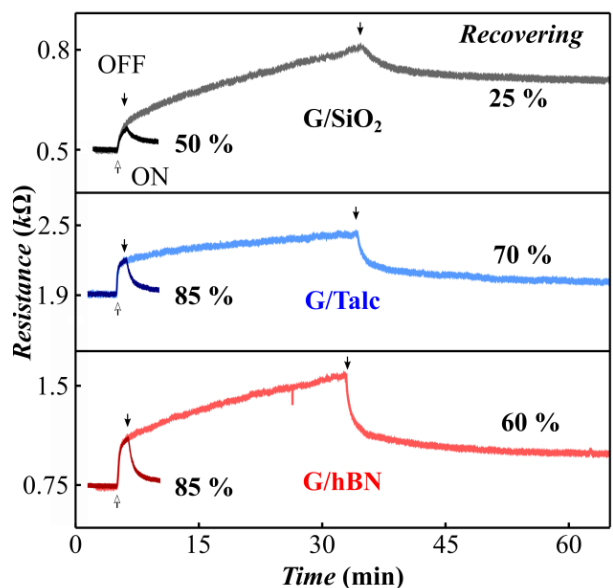

Figure 5: Graphene resistance as a function of time at 300 K under 10 % of NH$_3$ in N$_2$ as diluting gas. The results for the G/SiO$_2$ sensor is shown by the black curves, the G/talc sensor by the blue curves, and the G/hBN sensor by the red curves. The gate voltages applied are: -60 V, 2 V and -3 V for G/SiO$_2$ G/Talc and G/hBN devices, respectively. The arrows indicate when the NH$_3$ gas is turned on (up) or off (down), and the recovering percentages for each condition are also shown.

Now, let us emphasize that our findings show that graphene sensors with small G/substrate distances are superior sensors than G/SiO$_2$ devices. The reason behind is that they provide more suitable figure of merits for sensing applications such as: larger resistance response and faster recovering time, while G/SiO$_2$ sensors only demonstrate larger charge transfer with longer recovering time. Another important aspect analyzed here, is the capability of such sensors to operate in a less controlled environment (under the influence of humidity, for instance). We show that G/hBN devices are in fact superior than G/SiO$_2$ devices since their performance is weakly changed under low humidity conditions (dry air as carrier gas) or high humidity conditions (wet air: 80 % of relative humidity). We performed measurements of the resistance of the G/hBN device at 300 K under 10 % of NH$_3$ in different gas environment. As we show in the Fig. 6(a), the resistance variation increases from ΔR ~ 250 Ω under Ar + NH$_3$ (black curve), up to ΔR ~ 280 Ω and ΔR ~ 335 Ω under dry air + NH$_3$ (red curve) and wet air + NH$_3$ (blue curve), respectively. For such measurements, wet air environment is obtained by flowing dry air through a water bubbler bottle (as illustrated in Fig 1(a)) keeping a constant relativity humidity (~ 80 %) and we fixed the back gate at -2.25 V. By analyzing Fig. 6(a), one can observe that the presence of oxygen in the dry air and water slightly enhances the resistance response from NH$_3$ exposure. Note that similar behavior is observed for short time of NH$_3$ exposure (see Fig. S4 of supplementary material). Next, we would like to show that resistance change under NH$_3$ interaction is not only due to charge transferred from ammonia to graphene. In Fig. 6b we show the amount of charge per unit of area transferred from ammonia to graphene under interaction in different gas environment: Argon, dry air and wet air. Note that similar behavior is observed in dry air or Argon. However, measurements in wet air show a smaller charge transfer, even though there is a larger change in the resistance. It is important to point out that when there is a charge transfer process between graphene and an adsorbed molecules, the adsorbed molecule may become a source of Coulomb charge scattering [49,50]. Thus, the graphene conductivity ($\sigma = en\mu$) depends on two parameters $n$ (density of charge) and $\mu$ (charge mobility) been strongly affected by both factors. Our experiments suggest that in high quality G/hBN stronger charge scattering is occurring due to the interaction with ammonia under high humidity. The specific mechanism behind this phenomenon is not in the scope of this work and more detailed study must be performed to further understand it. Also, it is also important to mention that the electrical properties of G/SiO$_2$ and G/talc devices are significant affected by humidity or dry air flow indicating that such oxide surface interacts with H$_2$O and O$_2$ molecules [10,51,52]. Hence, these interactions can generate misleading results during NH$_3$ sensing (see also supplementary material for details).

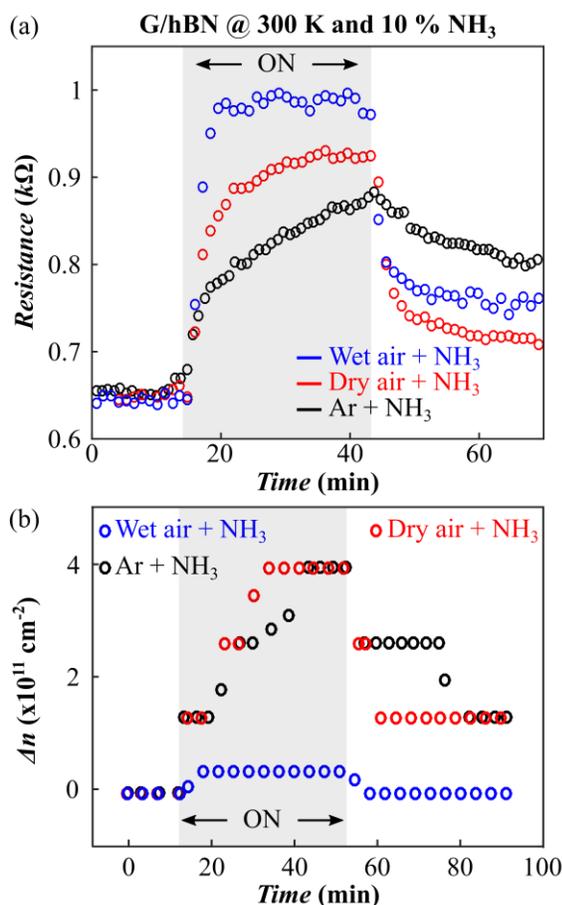

Figure 6: G/hBN sensor performance under 10 % NH$_3$ at different air environment: Ar + NH$_3$ (black data); dry air + NH$_3$ (red data); wet air + NH$_3$ (blue data). (a) Graphene resistance as a function of time and (b) Time evolution of the charge transferred per unit of area to graphene. In both measurements the temperature is set at 300 K and by diverting dry air through a water

bubbler bottle we set the relative humidity at 80 %. In the measurements in pure Argon or dry air the relative humidity is set at ~ 0 %. The gray regions show the time interval when the ammonia gas is turned ON into the system.

Finally, the charge transfer characteristics of graphene devices under 10 % of NH$_3$ exposure are also investigated at different operating temperatures, ranging from 300 K up to 450 K, as shown in Fig. 7 (a). One notices that the charge transfer decreases with increasing temperature, independently of the substrate, and that the G/SiO$_2$ sensor exhibits the highest charge transfer for all temperatures considered. The reduction in the charge transfer is also observed in our molecular dynamics simulations for the ammonia-graphene system at finite temperatures, $T$ = 300 K and 600 K, as shown in Fig. 7 (b). In the inset we present a snapshot of the 600 K simulation after 1 ps. In both cases, the molecule depicts a diffusive behavior on graphene, without desorption within the simulation times considered (1.972 ps at 300 K and 2.382 ps at 600 K). Although the molecule remains adsorbed at both temperatures, we observed a progressive reduction of the charge transfer with temperature. Such reduction has also been observed [14] and is explained based on the fact that thermal fluctuations facilitate detachment of NH$_3$ molecules from graphene surface. Additionally, we observe that the values of the time constants ($\xi_1$ and $\xi_2$) do not change considerably in going from a lower temperature to a higher temperature, indicating that the temperatures only act by reducing the amount of charge transferred from NH$_3$ to graphene devices.

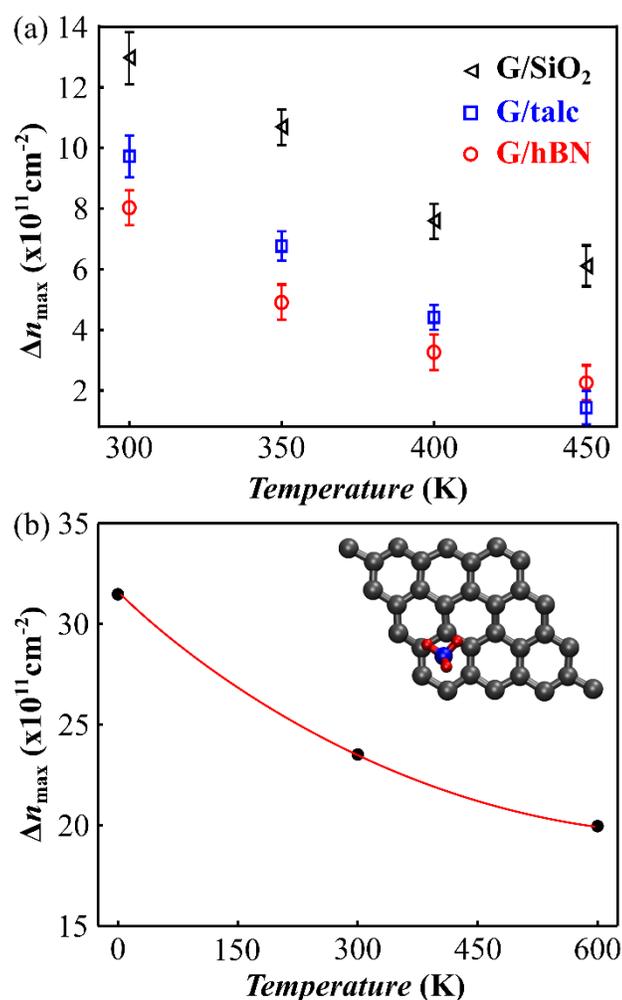

Figure 7: (a) Maximum of charge transferred ($\Delta n_{max}$) to graphene devices for a 10 % $NH_3$ concentration in $N_2$ as diluting gas, as a function of temperature for all devices: G/$SiO_2$ (black triangles), G/talc (blue squares) and G/hBN (red circles). (b) Molecular dynamics results of $NH_3$ molecules adsorbed to the graphene layer, which depicts the charge transferred to graphene as a function of temperature. The black dots are the first-principles results, and the red curve is a parabolic interpolation. Inset in (b) depicts a snapshot at $t = 1.0$ ps and $T = 600$ K of a $NH_3$ molecule atop of graphene sheet.

An interesting feature shown in Fig. 7 (a) is the difference between the charges transferred for each substrate. The charge transferred from $NH_3$ molecules to G/$SiO_2$ device is almost twice as large as that of the G/hBN device, and considerably superior than in G/talc at all temperatures. We propose that this effect results from the differences on the G/substrate distances. Graphene devices with larger distances between graphene and the substrate would accept a larger amount of charge from ammonia molecules, corroborating with the idea that the molecules are allowed to diffuse and interact with graphene at both sides of the graphene sheet. This agrees with our theoretical results (see supplementary material) which demonstrate that the charge transfer is essentially additive: for instance, one ammonia molecule transfers 0.029 *e* to graphene, where *e* is the electron charge; two molecules transfer 0.058 *e*, and so on, independent on the graphene "side" that the molecules are adsorbed. Therefore, ammonia molecules would donate electrons to graphene as soon as adsorbed to its surface. Note that the "best scenario" for charge transfer can be exemplified by the suspended case: since both sides of graphene are free to interact with molecules, the total of charge transference would occur when $\xi_1/\xi_2 = 1$. For graphene devices supported onto substrates, the smaller is the distance of separation between graphene and substrate, the more hampered is the molecular diffusion ($\xi_1/\xi_2 \ll 1$). Consequently, less charge would be transferred to the bottom side, resulting in a reduction of the total charge transferred, corroborating with our results. Therefore, our findings confirm that the G/substrate distance plays an important role on NH$_3$ graphene-sensor characteristics such as change transfer, resistance response and recovering. Additionally, it confirms that in order to develop new graphene sensors one must to consider the substrate engineering as a fundamental step during the devices fabrication [27–30].

## 4. CONCLUSION

In summary, we compared the performance of graphene-ammonia gas sensor with graphene on top of different substrates such as SiO$_2$, talc, and hBN from 300 K up to 450 K. We demonstrated that molecular doping of graphene from ammonia molecules is strongly dependent on the distance between graphene and substrate. We also compare our experimental data with DFT results, confirming that the electron charge transfer is higher when both sides of graphene sheet interact with ammonia molecules. Moreover, we show that G/hBN devices exhibit a faster recovery time and higher resistance response in comparison to G/SiO$_2$ devices and are slightly affected by changes in the air environment (dry or wet). Consequently, based on our findings we believe that the substrate engineering is a crucial point for development of graphene-based sensors and electronic devices, opening additional routes for faster devices with low power consumption.

### ACKNOWLEDGMENTS

This work was supported by CAPES, Fapemig (Rede 2D), CNPq and INCT/Nanomateriais de Carbono. The authors are thankful to Lab Nano at UFMG for allowing the use of atomic force microscopy. K.W. and T.T. acknowledge support from the Elemental Strategy Initiative conducted by the MEXT, Japan and JSPS KAKENHI Grant Numbers JP15K21722.

# Supplementary Material

# Enhancing the response of $NH_3$ graphene-sensors by using devices with different graphene-substrate distances

A. R. Cadore,[1,*)] E. Mania,[1] A. B. Alencar,[2] N. P. Rezende,[1] S. de Oliveira,[1] K. Watanabe,[3] T. Taniguchi,[3] H. Chacham,[1] L. C. Campos,[1] R. G. Lacerda[1]

[1]*Departamento de Física, Universidade Federal de Minas Gerais, Belo Horizonte, 30123-970, Brasil*

[2]*Instituto de Engenharia, Ciência e Tecnologia, Universidade Federal dos Vales do Jequitinhonha e Mucuri, Janaúba, 39440-000, Brasil*

[3]*National Institute for Materials Science, Namiki, 305-0044, Japan*

[*)] *Electronic mail: alissoncadore@gmail.com*

## 1. Transfer curve under $NH_3$ exposure for each device

The transfer curve ($R$ x $V_G$) of graphene devices is investigated at 300 K and exposed under 60 minutes to 10 % $NH_3$ gas diluted in ultrahigh pure $N_2$, keeping the total gas flow constant at 500 sccm. Figs. S1a-c show the time evolution of the charge neutrality point (CNP) for $NH_3$ exposure for a $G/SiO_2$, G/hBN and G/Talc devices, respectively. Note that we consider $t_0$ as the last curve in pure carrier gas, and $t_f$ as the curve after 60 min under $NH_3$ exposure. By sweeping the gate voltage, we measure the shift of the charge neutrality point (CNP) in the transfer curve ($\Delta V^{CNP}$) in response to the adsorption of $NH_3$. In all cases, the CNP shifts towards negative gate voltages: for $G/SiO_2$ it is positioned initially at around -5 V and over the time it moves up to -35 V; while for G/hBN it is initially at around -0.2 V, moving up to -1.1 V, and G/Talc it was at around 7 V, moving down to 5.9 V. The shift of the CNP towards negative gate voltages, as expected, shows clearly that $NH_3$ molecules act as electron donor, independently of the substrate, [1–5]. Moreover, notice that during the interaction, not solely the position of the CNP changes, but also there is a variation in the shape of the graphene transfer curve. Such result depicts that the ammonia molecules modify the charge scattering process, reducing the graphene electronic mobility similarly for electron and hole carriers, as can be seen in Figs. S1d-f for a $G/SiO_2$, G/hBN and G/Talc device, respectively.

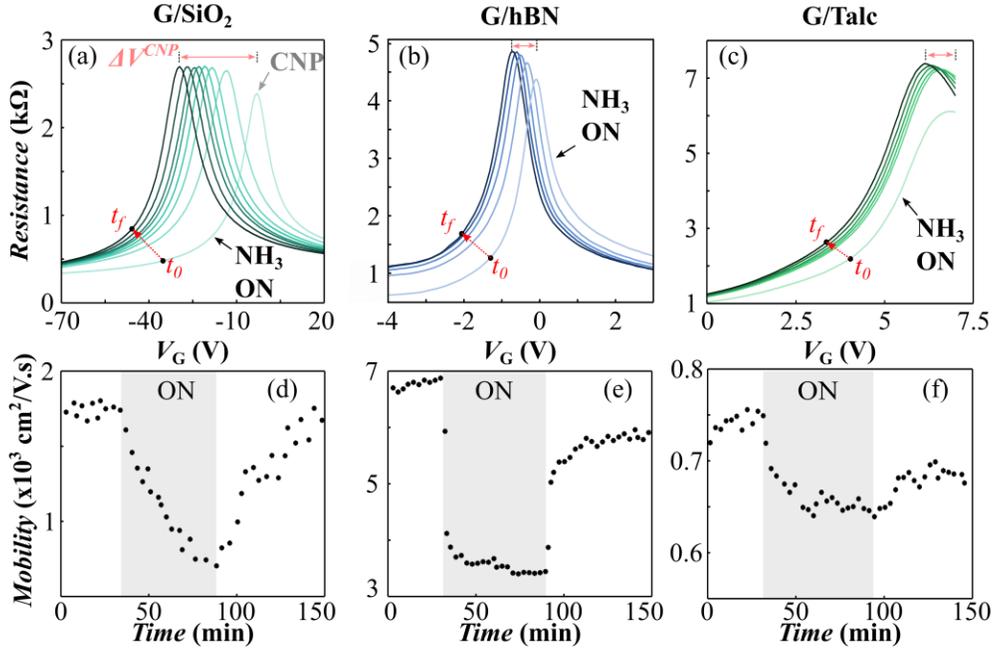

**Figure S1:** Resistance as a function of gate voltage ($V_G$) for G/SiO$_2$ (a), G/hBN (b) and G/Talc (c) devices at 300 K in N$_2$ as diluting gas. First, we set initial conditions by applying a flow of ultra-pure N$_2$, then we insert 10 % of ultra-pure NH$_3$, keeping the total gas flow constant. The black arrows point at the last measurement ($t_0$) before 10 % of ammonia is turned ON in the system. $t_f$ is defined after 60 min under ammonia exposure. $\Delta V^{CNP}$ illustrates the voltage difference between both CNP before and after NH$_3$. The maximum mobility as a function of exposure time to NH$_3$ gas for G/SiO$_2$ (d), G/hBN (e) and G/Talc (f) devices.

From the transfer curves showed in Figure S1 we calculate the absolute value of the density of charge transferred from ammonia to graphene ($\Delta n$), as we present in Figures 3a-c of the main text. $\Delta n$ is calculated from the shift of the gate voltage of the CNP before and after the ammonia exposure ($\Delta V^{CNP}$). Using the well-known parallel plate capacitor formula $\Delta n = (\varepsilon \varepsilon_0 \Delta V^{CNP})/ed$ [2,6]. In this equation, $\varepsilon$ is the dielectric constant of the substrate (SiO$_2$ = talc = hBN = 3.9), $\varepsilon_0$ is the vacuum permittivity, $d$ is the dielectric thickness (SiO$_2$ = 285 nm, talc = 18 nm or hBN = 20 nm), and $e$ is the electron charge. Note that a similar approach is used to calculate the amount of charge transferred to graphene under different air environment in Figure 6b. For such analyze the G/hBN device was fabricated with 22 nm hBN thick.

## 2. Analyzes of air environment in graphene devices and ammonia sensing properties

First of all, we tested the influence of dry air in our devices in comparison to Argon atmosphere as presented in Figure S2. In dry air the G/SiO$_2$ sensor is considerably depleted of electrons becoming a material with hole-type as charge carries. Such depletion can be described by the transfer of electrons from the graphene layer to the O$_2$ molecules presented in the dry air [7,8]. On the other hand, G/hBN devices are very stable in both Ar and dry air atmospheres. Comparing both substrates, we can conclude that G/hBN devices are more stable than G/SiO$_2$ sensors due to changes in the environment.

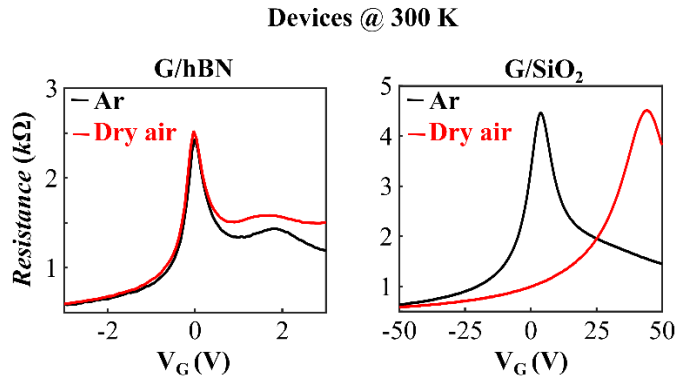

**Figure S2**: Comparation between G/hBN (left) and G/SiO$_2$ (right) devices under ultra-pure argon (black curves) and ultra-pure dry air atmosphere (red curves).

Secondly, we tested the influence of humidity in our devices in comparison with Argon atmosphere. From Figure S3, one can clearly see that, without application of gate potentials, the G/SiO$_2$ and G/talc devices, just after the process of fabrication (black curves in the figures), are highly depleted of electrons with hole-like charge carries. For instance, the CNP are located at a back-gate voltage ($V_G$) $V_G > 50$ V and $V_G > 6$ V for G/SiO$_2$ and G/talc, respectively, while the G/hBN device is located close to $V_G = 0$ V indicating no charge transfer due to the environment. Such results show that solely G/hBN devices are not affected by elements present in the environment of the laboratory due to most likely from humidity [7–9].

After the standard conditioning at 200 °C during overnight (12h) in Argon atmosphere, one can also observe that the transfer curves from both G/SiO$_2$ and G/talc devices decreased considerably their initial doping (red curves in the figures S3 (a)-(c)), indicating that this doping was, in fact, most probably caused by H$_2$O molecules adsorbed on graphene surface [7–9]. One can note that G/hBN transfer curve have just slightly changed after conditioning. Now, we show results for measurements in dry air (green), and for 80 % of humidity (blue) inside the chamber. Note that this humidity level was intentionally created. One can see that in all conditions the G/hBN device maintain a constant behavior at the CNP, independent of the air environment. There is only a small change in the graphene resistance at high doping level (asymmetry between electrons and holes) and such variation can be associate with changes in the Fermi level pinning at the metal-graphene interface caused by different gases [10]. Conversely, G/SiO$_2$ sensors are strongly affected by the atmosphere, demonstrating that sensors using graphene on hBN substrate are more stable than on SiO$_2$. Unfortunately, we were not able to perform the same set of measurements on G/talc devices, but we expect a similar doping as observed for SiO$_2$, because the initial conditioning already shows a strong humidity influence.

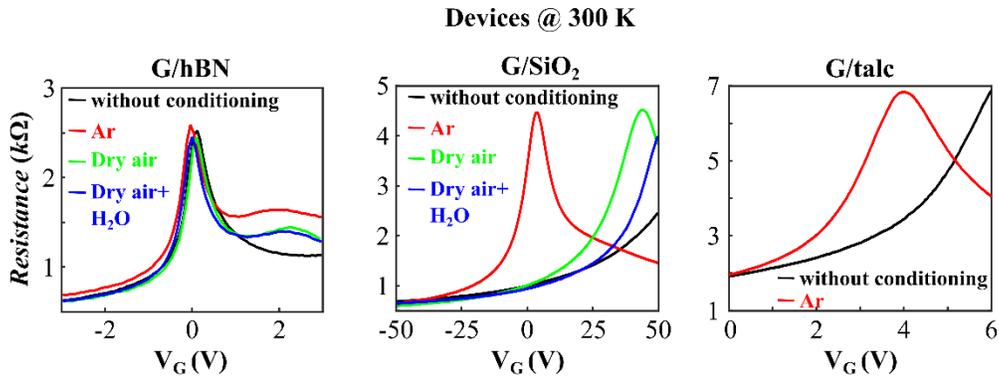

**Figure S3:** Comparation between G/hBN (left), G/SiO$_2$ (center) and G/talc (right) devices under different air environments: without conditioning (black curves); pure argon (red curves); pure dry air (green curves); wet air (blue curves).

## 3. Influence of the air environment and humidity on the NH$_3$ sensing properties under short time exposure time

As discussed in the main text, the sensing properties are slightly modified by either dry air environment or wet air when compared to a pure argon atmosphere for long time of exposure. Here, for instance, we performed measurements under pulses (short time exposure) of 10 % of NH$_3$ in different gas environment for the same G/hBN device. One can observe the results are quite similar to the long exposure time presented in Fig. 6(a). Besides, in all cases the sensor shows good reproducibility and recovering time (reaching ~ 75 % after 5 min of degassing). Similarly, as presented for long time of exposure in the main text, we fixed the back gate at -2.25 V and relativity humidity at ~ 80 %.

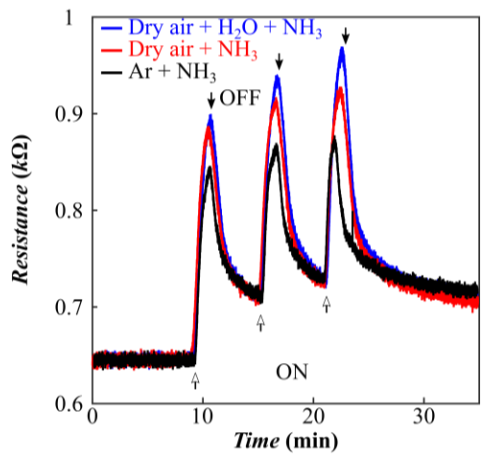

**Figure S4:** Graphene resistance as a function of time for pulses - 1 min ON (arrows up) and 5 min OFF (arrows down) – under 10 % NH$_3$ exposure for the G/hBN sensor at different air environment: Ar + NH$_3$ (black data); dry air + NH$_3$ (red data); dry air + H$_2$O + NH$_3$ (blue data). In all measurements the temperature is set at 300 K and by diverting dry air through a water bubbler bottle we set the relative humidity at 80 %. In the measurements in pure Argon or dry air the relative humidity is set at ~ 0 %.

## 4. DFT Analyses

The first principle calculations were performed based on the SIESTA [11] implementation for density functional theory (DFT) [12,13]. We made use of a van der Waals (VDW) [14,15] functional for the exchange-correlation potential. We employed norm-conserving Troullier-Martins pseudopotentials [16] in the Kleinman-Bylander factorized form [17], and double-zeta basis set augmented by polarization functions (DZP). A real space mesh was used with a cutoff of 450Ry. All geometries were optimized so that the maximum force component on any atom was less than 10meV/atom. Two distinct cases of ammonia adsorption were considered. In the first case, shown in Fig. S5a, an ammonia molecule is deposited atop a suspended graphene sheet, modelled by a periodic repetition of a 4x4 graphene unit cell. In the second case, shown in Fig. S5b, the molecule is deposited atom a graphene/talc heterostructure, modelled by a periodic repetition of a 2x2 talc unit cell atop a 4x4 graphene unit cell. The two most stable configurations of the ammonia molecule [18] atop graphene are also shown in Figs. S5a and S5b. In Fig. S5a the molecule is in the up-center (uc) configuration, and, in Fig. S5b, in the down-center (dc) configuration. We find that the ammonia molecule in the uc configuration is a donor both atop graphene and atop graphene/talc, donating 0.029 $e$, where $e$ is the electron charge, in both cases. In the dc configuration the molecule is an acceptor, receiving 0.012 $e$ in the first case and 0.011 $e$ in the second case. We also considered the case where talc is already doped by an acceptor impurity (substitutional Al on a Si site). In this case, the uc ammonia molecule is still a donor, donating 0.029 $e$, while the dc molecule is nearly non-dopant. Note that all the calculations described above were performed at zero temperature, with the geometries optimized. However, we also considered molecular dynamics simulations for the ammonia-graphene system at finite temperatures as discussed in the main text.

The doping of carbon nanomaterials by adsorbed species has been investigated for several years, starting with the electron transfer from the carbon nanotube to the adsorbed oxygen [19,20] and the electron transfer to the carbon nanotubes from adsorbed ammonia molecules [9,21]. In the case of oxygen doping, DFT calculations by the Steven Louie and Marvin Cohen groups [19] have predicted that a carbon nanotube would be positively charged due to the electron transfer to the adsorbed oxygen molecule. This was confirmed experimentally [20] and the same process is observed in graphene devices [9]. In the case of ammonia interaction, DFT calculations have predicted that carbon nanotubes to be negative charged due to electron transfer from an adsorbed ammonia molecule, which was also confirmed experimentally [20]. In the present work, both our DFT calculations and measurements indicate the negative charge transfer to graphene by adsorbed ammonia molecules.

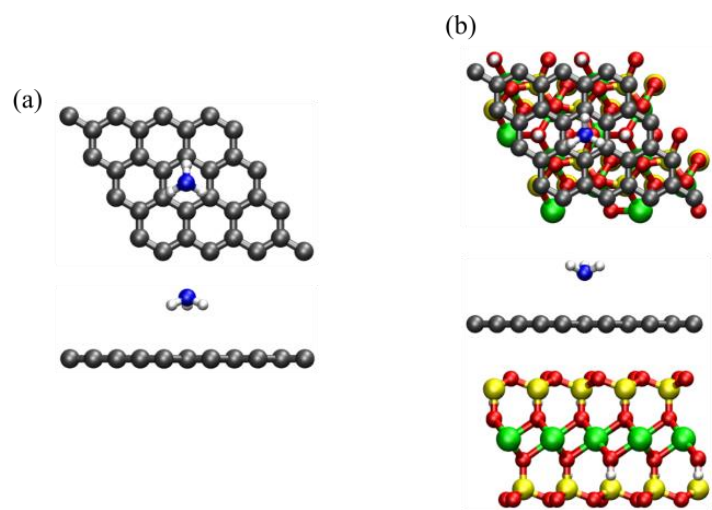

**Figure S5**: First-principles results at zero temperature for the adsorbed ammonia molecule (a) on graphene in the up-center configuration and (b) on the graphene/talc heterostructure in the down-center configuration.